# Liquid Metal Transformers

**Lei Sheng[1,3], Jie Zhang[1,3], Jing Liu[1,2*]**

1. Department of Biomedical Engineering, School of Medicine,
Tsinghua University, Beijing, China

2. Beijing Key Lab of CryoBiomedical Engineering and Key Lab of Cryogenics,
Technical Institute of Physics and Chemistry,
Chinese Academy of Sciences, Beijing, China

3. These authors contributed equally to this work.

***Address for correspondence:**

Dr. Jing Liu

Department of Biomedical Engineering,

School of Medicine,

Tsinghua University,

Beijing 100084, China

E-mail address: jliubme@tsinghua.edu.cn

Tel. +86-10-62794896

Fax: +86-10-82543767

**Abstract:**

The room temperature liquid metal is quickly emerging as an important functional material in a variety of areas like chip cooling, 3D printing or printed electronics etc. With diverse capabilities in electrical, thermal and flowing behaviors, such fluid owns many intriguing properties that had never been anticipated before. Here, we show a group of unconventional phenomena occurring on the liquid metal objects. Through applying electrical field on the liquid metals immersed in water, a series of complex transformation behaviors such as self-assembling of a sheet of liquid metal film into a single sphere, quick mergences of separate metal droplets, controlled self-rotation and planar locomotion of liquid metal objects can be realized. Meanwhile, it was also found that two accompanying water vortexes were induced and reliably swirled near the rotating liquid metal sphere. Further, effects of the shape, size, voltage, orientation and geometries of the electrodes to control the liquid metal transformers were clarified. Such events are hard to achieve otherwise on rigid metal or conventional liquid spheres. This finding has both fundamental and practical significances which suggest a generalized way of making smart soft machine, collecting discrete metal fluids, as well as flexibly manipulating liquid metal objects including accompanying devices.

# 1. Introduction

In nature, designing objects that can flexibly transform between different configurations and freely move via a controllable way to perform desired tasks has long been a dream among diverse scientific and technological areas ranging from biology to physics. Tremendous efforts have therefore been made to explore the motion of artificial swimmers [1], through strategies like electrodeposition [2] or electropolymerization to grow random [3] or directional [4] structures of conducting material between two electrodes. To avoid use of the chemical fuel, alternative driving approaches were also tried, such as light [5], ultrasonic propulsion [6], magnetic force [7], biomimetic propulsion [8] as well as electricity [9] etc. Among the many typical methods, electric field-driving is especially convenient for practical purpose. In this side, electrokinetic effects have been found to occur on particles when exposed to an electric field which have also been adopted to manipulate target objects [10]. Generally, if an electric field applies on an interface separating two immiscible liquids, it would undergo a jump due to transition of physical properties from one medium to another. One consequence of such field discontinuity is the presence of an electric stress at the interface. In the case of a suspended droplet placed in an otherwise uniform electric field, the curvature of the interface usually creates surface gradients of electric field and stress that are likely to deform the droplet. However, so far, most of the conductive droplets ever tested still fall in the category of a conventional liquid.

Recently, the room temperature liquid metals were found to have unique virtues in a wide variety of important areas such as chip cooling, 3D printing or printed electronics due to owing outstanding diverse physical capabilities, such as high conductivity as well as easy mobility etc. Mercury is a well-known room temperature liquid metal. Unfortunately, its toxicity poses a serious safety concern for the widespread applications. As an alternative, a class of room temperature liquid metal and its alloys were found to own very interesting property that may be advantageous as materials that are movable in a controllable way [11]. Such liquid metal has a broad temperature range of liquid phase with a melting point at 10.35 ℃. And they are generally chemically stable and do not react with water at around room temperature. A series of previous studies have proven that such alloy is safe for humans in many normal occasions. Particularly, the high conductivity of the alloy is up to $3.1*10^6$ $Sm^{-1}$ [12], which is several orders higher in magnitude than that of non-metallic materials and comparable with many other common metallic materials. Therefore, the liquid metal spheres or other manifestations formed in water can naturally function as active electronic junctions. With such intrinsically existing conductive liquid metal immersed in water, it is expected to achieve various electrically controlled behaviors which had never been observed before. Through years' continuous working on liquid metal, here we occasionally found a group of unconventional transformation phenomena thus happen which may have generalized importance either for fundamental science or application purpose. It was demonstrated that when applying electricity on the target liquid metal GaInSn alloy immersed in water, a series of transformation behaviors such as self-assembling of a sheet of liquid metal film into a single sphere, automatic mergences of any contacting liquid metal droplets, planar locomotion and rotation of the liquid metal can be easily realized. Meanwhile, we also found that the rotation of the metal sphere would induce accompanying vortexes in the surrounding water. The reliable running of the liquid metal transformers suggests that it is very practical to make a soft machine along this way in the near future.

## 2. Materials and Methods

The present experiments on the electrically controlled liquid metal transformation phenomena were carried out on the GaInSn alloy in water. GaInSn alloy was prepared from gallium, indium and tin metals with purity of 99.99 percent. Such raw materials with a volume ratio of 67:20.5:12.5 were added into the beaker and heated at 100 ℃. Then a magnetic stirrer was utilized to stir the mixture after they were all melted to achieve uniform mixing. The experiments were performed using a setup consisting of Petri dish, GaInSn alloy, electric cords with copper wire and power supply, respectively. All equipment is presented in Figure 1(a-e) and the apparatus is assembled as shown in Figure 1(f). Alternatively, a high-speed camera Canon XF305 was adopted for shooting the experimental phenomena.

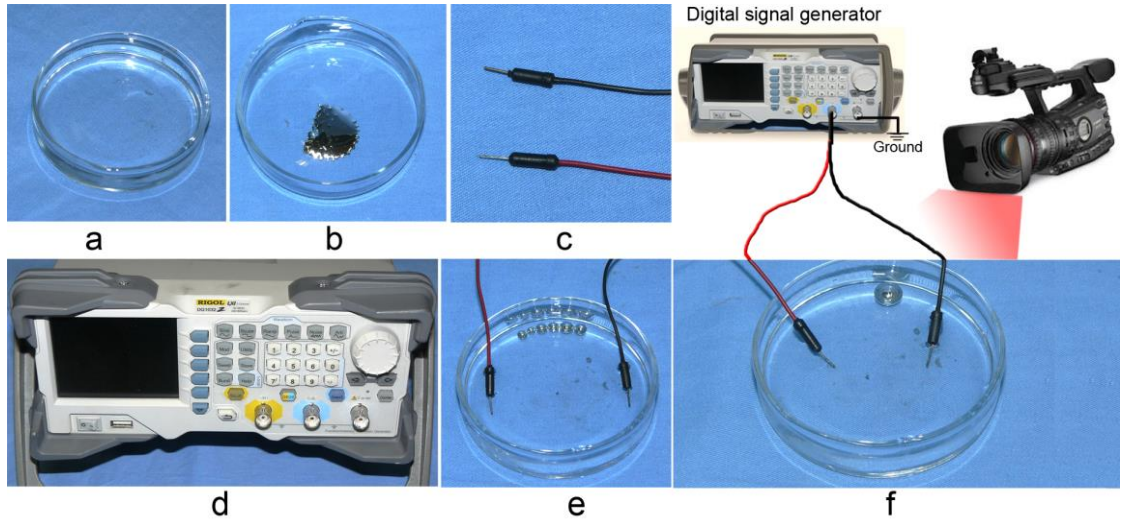

**Figure 1| Equipment and experimental diagram on testing liquid metal transformation phenomena.** (a) Water and Petri dish. (b) A sheet of liquid metal splashed in water. (c) Electric cords with copper wire. (d) Power supply. (e) Liquid metal spheres in water with electric cords. (f) Close-up of the apparatus with camera.

## 3. Experimental Results

### 3.1 Transformation and mergence of liquid metal objects

Following the procedures as outlined in Figure 2a, a drop of liquid metal was added onto the surface of the plastic base to form a flat oval which is about 2 cm in its long axis (Figure 2b). Then the same amount of water was dropped on the liquid metal to contain it inside. The anode of the copper wire was immersed in water, and the cathode of the copper wire was attached to the liquid metal. If switching on the electricity, the shape of the liquid metal would quickly transform from its original flattened configuration into a spherical one (Figure 2b), indicating significant variations of the interfacial tension applied on the liquid metal. Here direct current was applied with a voltage set as 12V. For a much larger sheet of liquid metal film covering the whole dish, similar phenomena were also observed. Typical sequences for this kind of liquid metal transformation are shown in Figure 2c.

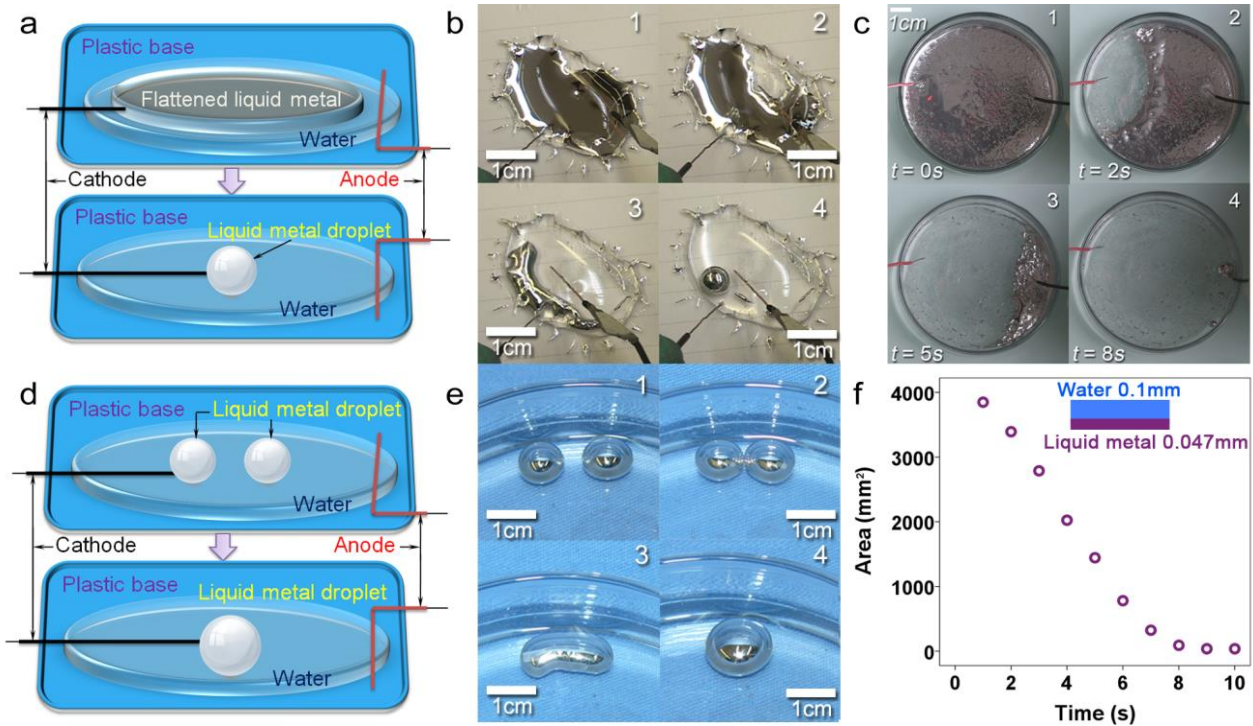

**Figure 2| The electric field-induced transformation of liquid metal objects in water.** (a) Schematic for electric field-induced transformation from a pool of liquid metal into a sphere. (b) Liquid metal shape transformation from original flattened state 1 to intermediate states 2 and 3 until finally spherical shape 4. (c) Sequential transformation for a circular sheet of liquid metal film to change from its initial flattened state 1 to intermediate states 2 and 3 until finally spherical shape 4, the original area of liquid metal film is 3848mm$^2$, while the thickness for liquid metal and water layers is 0.047mm and 0.1mm, respectively. (d) Schematic for electric field-induced mergence from two separate spheres into a single one. (e) Two separate droplets combining into a fused sphere from state 1 to 4 when put to contact together. (f) The transient area variation for liquid metal film to change from its original flattened shape into a sphere.

According to several former theoretical predictions [13], it can be understood that the liquid metal object in water structural form is mainly dominated by the surface tension. Such effect would exhibit a number of unique properties, including very small contact area with surrounding surfaces leading to low friction rolling, super-hydrophobic interactions with other fluids and the ability to be split or fused together with self-healing encapsulation layers. With such effect in mind, our additional experiments further discovered that splited liquid metals could be easily collected through simply applying electricity on them (Figure 2d). To test the ability to merge different liquid metal objects in the glass dish, we placed two dispersed liquid metal objects in close proximity and slightly contact them together. Then a quick mergence between the metal droplets happens. Figure 2e presents the original separate droplets, the application of a spatula to contact them together, and the resulting fused sphere. Clearly, the final sphere was relatively large and appears quite spherical in shape.

For more specific information on the transient liquid metal shape transformation, a quantitative analysis was performed on Figure 2c, where the original area of liquid metal shape is 3848mm2, and the thickness for the liquid metal or water layers is 0.047mm and 0.1mm respectively. The result is depicted in Figure 2f. It can be noticed that, the total time for the liquid metal film in Figure 2c to finish its transformation from the initial flattened circular shape into a sphere is about 10s. When this liquid metal sphere was cut apart using a fine copper wire, it would split into two small spheres and remain spherical in shape. Such mechanism offers an extremely useful way towards making soft transformers and also for future liquid metal recycling.

## 3.2 Rotation of liquid metal sphere and its induced water vortexes

Except for the above electricity induced configuration transformations, additional unconventional phenomena were also found to happen on the present two-fluid system made of liquid metal and water. To test the effect, the liquid metal droplet was immersed in the water. Then two electrodes were placed at the relative positions as shown in Figure 3a. When switching on the electricity, an automatic rotation of the liquid metal sphere was observed (Figure 3b). Meanwhile, it was also found that there simultaneously occurred two accompanying vortexes in the nearby water, which kept swirling at the cathode side. Here the arrows on the picture indicate direction of the vortex motion in the water. The swirling of the water vortexes near the liquid metal sphere can be visualized by using suspension particles generated in advance through electrolytic effect on the electrode (Figure 3c). Clearly, the rapid rotation of both liquid metal sphere and water vortexes runs rather reliably. Such behavior is hard to achieve otherwise for a rigid metal or conventional conductive sphere especially when no magnetic fields were involved. If changing the relative orientations of the two electrodes, similar phenomena were also observed. Plots of gray scale along the horizontal lines at the middle height of swirls labeled as a1 and b1 in Figure 3c were presented in Figure 3d. The symmetrical curve in gray scale indicates existence of the two swirls at a1 and b1. The reasons for causing these abnormal fluidic behaviors can all be attributed to the liquid metal's diverse properties like highly conductive however flowable features. Overall, the convective motions were driven by the interfacial tension gradients of the liquid metal sphere.

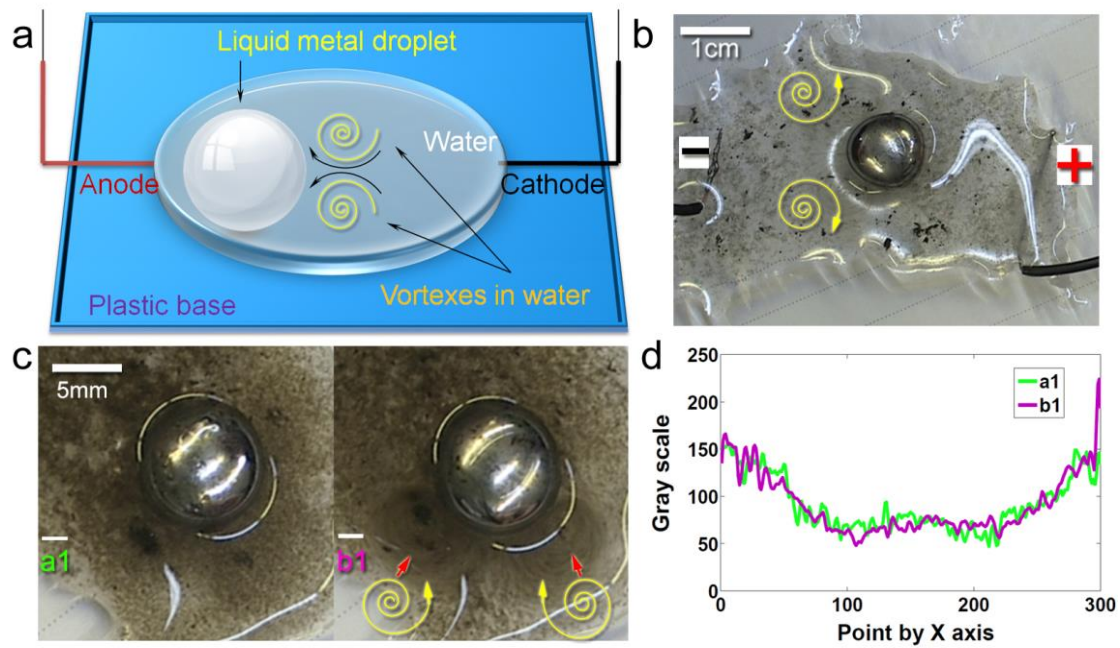

**Figure 3| The electric field-induced rotation of liquid metal sphere and accompanying vortexes in water.** (a) Schematic for electric field-induced

rotation. (b) Liquid metal sphere rotation and its induced accompanying vortexes in water, where yellow arrows indicate direction of the swirls close to the cathode. (c) Another two cases of liquid metal sphere rotation and vortexes in water, where the swirling of the water near the liquid metal electrode could be visualized by using suspension solid particles there. (d) Plots of gray scale along the horizontal line at the middle height of the swirls labeled as a1 and b1 in Figure 3c, the symmetrical curves of the gray scale indicate existence the two swirls.

**3.3 Planar locomotion of liquid metal objects**

Through a series of conceptual experiments, we further clarified that nearly every kind of complex flow behavior can be induced on the liquid metal object. Specifically, an electrically controlled directional locomotion for the liquid metal sphere can be realized. For this purpose, a liquid metal sphere was immersed in a water channel pre-made on the plastic plate. The electric field was imposed by applying a 12 V DC between two electric cords which were vertically placed and separated by about 8 cm (Figure 4a). The DC voltages are generated by a signal generator. When the electricity was switched on, several forces were induced as depicted by Figure 4b and their imbalance resulted in a directional locomotion, as sequentially shown in Figure 4c. The driving forces included the surface tension gradient force induced by the electric field and the rotational force for water. To achieve a directional or reciprocal motion, these forces overcame the retardation effects including the viscous friction between the droplet and its surrounding electrolyte as well as the frictional force between the droplet and the surface of the substrate, resulting in droplet locomotion. The liquid metal sphere moved in an accelerated way towards the anode with an average speed of approximately 6 body length. A quantification on the transient locomotion of two different sized liquid metal droplets was shown in Figure 4d. Clearly, the larger diameter for the droplet, the quicker it moves. It is worth to note that no movement was observed when the applied voltage was lower than 12 V under current experimental conditions. The present study adds new findings to the former locomotion phenomenon as observed before on some conductive objects especially liquid marble coated with nanoparticles on the sphere surfaces [14-16]. In the electrochemically induced interesting chemical locomotion [16], although a planar movement of the conducting objects could be induced therein, the motion speed is still somewhat slow and the material is not as common as identified in the present study.

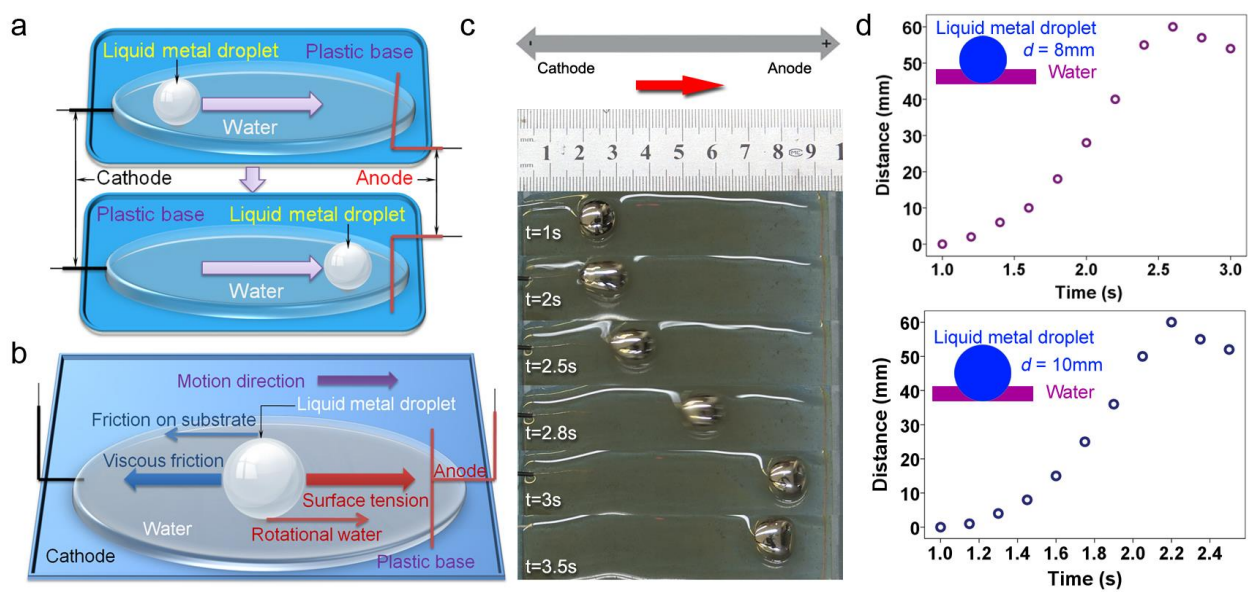

**Figure 4| Planar locomotion of liquid metal sphere induced by electric field.** (a) The diagram of electric field-induced planar locomotion. (b) Schematic of showing the forces affecting the motion of a liquid metal sphere in water induced by electric field. (c) The photograph of liquid metal sphere motion in water, sequential snapshots for liquid metal sphere moving in water when a DC voltage of 12V was applied. (d) The transient locomotion distance for two different sized liquid metal droplets.

## 4. Discussion

Overall, the most important fundamental discovery as achieved by the present work is that nearly every kind of transformation behaviors can be realized on the liquid metal objects. It opens a generalized way to make smart liquid machine. With easily controllable feature, such artificial transformer is expected to offer plenty of practical opportunities in the coming time. It is worth to mention that all such transformations were enabled by only very low electrical voltage. Therefore the scope of the applications thus involved can be extensive. Further, liquid metal and water are two materials quite common in nature. And the liquid metal alloy is safe for humans in normal occasions, owing to their excellent biological compatibility [17]. As a result, encapsulating them together with elastic materials may create potential to serve as artificial machine. A soft robot can thus be made from such liquid metal transformers. Meanwhile, it also posed promising prospects in the motion needed devices that can be implanted in human body.

Further, the choice of different forms of electrical fields allows occurrence of a variety of phenomena that are worthy of study. For example, a number of electrodes can be aligned in parallel, while the water and liquid metal can be packaged in earthworm-shaped structures. When the square wave voltage signal was applied, an earthworm-like soft robot might be driven to move. The materials as adopted in this work are moldable and flexible, and the fabrication process is simple. They are compatible with water-based or high humidity environments. Since biological systems generally have soft, curved and in some cases moving surfaces and tend to be operated by electronic current, the present soft, electronic current controlled mergence may find potential applications in recycling the liquid metal residues injected into biological body as medical electronics [17], soft-matter diodes with liquid metal electrodes [18], bio-embeddable smart particles and biomimetic devices. We attributed the mechanisms to cause such liquid metal transformations to the dynamic balance between the surface tension and the electronic force applied on the liquid metal surface. The present behavior belongs to certain kind of self-organization on macroscopic scale. Whether such mechanism can make expected transformation into cooperating rolls or cells in micro scale deserves further study.

## 5. Conclusion

In summary, electric field induced transformations of liquid metal in different styles were discovered. Along with rotation of the liquid metal sphere, accompanying water vortexes were also induced which run reliably. Owing to the diverse capabilities of the liquid metals in underwater electric field, it opens important opportunities for the practices of liquid metal recycling, soft machine manufacture, locomotion control of objects as well as moveable sensors, microfluidic valve, pump or more artificial robots. Such liquid metal transformers and locomotors could provide on demand use given specific designing. Importantly, a smart liquid metal machine could be extended to three dimensions when a spatial electrode configuration is adopted. Further, the phenomena of those complex transformation behaviors without gravity effect are also worth of pursuing in the near future.